\documentclass[12pt]{book}
\usepackage{epsfig}
\newcommand{\ba}{\begin{array}}
\newcommand{\be}{\begin{equation}}
\newcommand{\bea}{\begin{eqnarray}}
\newcommand{\bean}{\begin{eqnarray*}}
\newcommand{\ben}{\begin{enumerate}}
\newcommand{\bi}{\begin{itemize}}
\newcommand{\een}{\end{enumerate}}
\newcommand{\ei}{\end{itemize}}
\newcommand{\ea}{\end{array}}
\newcommand{\ee}{\end{equation}}
\newcommand{\eea}{\end{eqnarray}}
\newcommand{\eean}{\end{eqnarray*}}

\newcommand {\x}{${\bf x}$ }
\newcommand {\bb}{${\bf b}$ }
\newcommand {\A}{${\bf A}$ }

\begin{document}

\title{\bf Vector Space and Matrix Methods in \\ Signal and System Theory}
\author{{\it C. Sidney Burrus }\\
    Electrical and Computer Engineering Dept.\\
    Rice University, Houston, Tx 77005\\
    csb@rice.edu\\
    Open license under Creative Commons (cc-by-4.0)
    }

\maketitle
\tableofcontents             
\chapter{Introduction}
The tools, ideas, and insights from linear algebra, abstract algebra, and functional
analysis can be extremely useful to signal processing and system theory in various areas
of engineering, science, and social science including approximation, optimization, parameter 
identification, big data, etc.  Indeed, many important ideas can
be developed from the simple operator equation
\be
    {\bf A x = b}
\label{1}\ee
by considering it in a variety of ways.  If \x and ${\bf b}$ are
vectors from the same or, perhaps, different vector spaces and ${\bf A}$
is an operator, there are three interesting questions that can be asked
which provide a setting for a broad study.
\ben
\item Given \A and \x , find \bb. The analysis or operator problem or transform.
\item Given \A and \bb , find \x. The inverse  or control problem or deconvolution or design 
	or solving simultanious equations.
\item Given \x and \bb , find \A. The synthesis  or design problem or parameter identification.
\een
Much can be learned by studying each of these problems in some detail.  We
will generally look at the finite dimensional problem where (\ref{1}) can more 
easily be studied as a finite matrix multiplication \cite{strang0,strang7,moler5,Trefethen2}
\vspace{10pt}
\begin{equation}
 \left[ \begin{array}{ccccc}
        a_{11} & a_{12} & a_{13} & \cdots & a_{1N}\\
        a_{21} & a_{22} & a_{23} &  & \\
        a_{31} & a_{32} & a_{33} &  & \\
        \vdots & & & & \vdots \\
        a_{M1} &  &  & \cdots & a_{MN}
        \end{array} \right]
            \left[ \begin{array}{c}
            x_1\\
            x_2\\
            x_3\\
            \vdots\\
            x_N\\
                  \end{array} \right]
  =
    \left[ \begin{array}{c}
    b_1\\
    b_2\\
    b_3\\
    \vdots\\
    b_M\\
         \end{array} \right]
\label{1aa}\end{equation} \\
\vspace{20pt}
but will also try to indicate what the infinite dimensional case might be \cite{halmos,young,oden,moon}.  

An application
to signal theory is in \cite{franks}, to optimization \cite{dgl1}, and multiscale system
theory \cite{bnw}.  The inverse problem (number 2 above) is the basis
for a large study of pseudoinverses, approximation, optimization, filter design, and
many applications.  When used with the $l_2$ norm \cite{ls,bjorck} powerful
results can be optained analytically but used with other norms such as
$l_\infty$, $l_1$, $l_0$ (a pseudonorm), an even larger set of
problems can be posed and solved \cite{albert,israel}.

A development of vector space ideas for the purpose
of presenting wavelet representations is given in \cite{id1,bgwave}.
An interesting idea of \emph{unconditional bases} is given by Donoho \cite{dld3}.

\emph{Linear regression} analysis can be posed in the form of (\ref{1}) and (\ref{1aa}) where the ${\bf M}$
rows of ${\bf A}$ are the vectors of input data from ${\bf M}$ experiments, entries of
${\bf x}$ are the ${\bf N}$ weights for the ${\bf N}$ components of the inputs, and 
the ${\bf M}$ values of ${\bf b}$ are the outputs \cite{albert}.  This can be
used in machine learning problems \cite{bishop,haykin3}.  A problem similar to the
design or synthesis problem is that of parameter identification where a model of some
system is posed with unknown parameters.  Then experiments with known inputs and
measured outputs are run to identify these parameters.  Linear regression is also an
example of this \cite{albert,bishop}.

{\it Dynamic systems} are often modelled by ordinary differential equation where
${\bf b}$ is set to be the time derivative of ${\bf x}$ to give what are called the linear state 
equations:
\be
    {\bf \dot{x} = A x}
\label{1a}\ee
or for difference equations and discrete-time or digital signals,
\be
    {\bf x}(n+1) = {\bf A \, x}(n)
\label{1b}\ee
which are used in digital signal processing and the analysis of certain algorithms. 
State equations are useful in feedback control as well as in simulation of many
dynamical systems and the eigenvalues and other properties of the square matix ${\bf A}$ are
important indicators of the performance \cite{zadeh2,derusso}.

The ideas of similarity transformations, diagonalization, the eigenvalue
problem, Jordon normal form, singular value decomposition,  etc. from linear 
algebra \cite{strang,strang0,strang7,hefferon} are applicable to this problem.

Various areas in optimization and approximation use vector space math to 
great advantage \cite{dgl1,ls}.

%
This booklet is intended to point out relationships, interpretations, and tools in linear algebra,
matrix theory, and vector spaces that scientists and engineers might find useful. It is not a 
stand-alone linear algebra book.  Details, definitions, 
and formal proofs can be found in the references.  A very helpful source is Wikipedia.

There is a variety of software systems to both pose and solve linear algebra
problems.  A particularly powerful one is Matlab \cite{moler5} which is, in some ways,
the gold standard since it started years ago as a purely numerical matrix package.  But
there are others such as Octave, SciLab, LabVIEW, Mathematica, Maple, R, Python, etc.
\chapter{A Matrix Times a Vector}

In this chapter we consider the first problem posed in the introduction
\be
    {\bf A x = b}
\label{1c}\ee
where the matrix \A and vector \x are given and we want to interpret and give structure to
the calculation of the vector \bb.  Equation (\ref{1c}) has a variety of special
cases.  The matrix \A may be square or may be rectangular.  It may have
full column or row rank or it may not.  It may be symmetric or orthogonal
or non-singular or many other characteristics which would be interesting
properties as an operator.  If we view the vectors as signals and the
matrix as an operator or processor, there are two interesting interpretations.
\bi
\item The operation (\ref{1c}) is a change of basis or coordinates for a
fixed signal.  The signal stays the same, the basis (or frame) changes.
\item The operation (\ref{1c}) alters the characteristics of the signal 
(processes it) but within a fixed basis system.  The basis stays the same, the signal
changes.
\ei
An example of the first would be the discrete Fourier transform (DFT)
where one calculates frequency components of a signal which are
coordinates in a frequency space for a given signal.  
The definition of the DFT from \cite{csbcnxfft} can be written as a matrix-vector
operation by ${\bf c=Wx}$ which, for $w = e^{-j2\pi/N}$ and $N=4$, is
\vspace{.1in}
\begin{equation}
\left[ \begin{array}{c}
        c_0\\
        c_1\\
        c_2\\
        c_3
        \end{array}\right] =
    \left[ \begin{array}{cccc}
        w^0 & w^0 & w^0 & w^0 \\
        w^0 & w^1 & w^2 & w^3 \\
        w^0 & w^2 & w^4 & w^6 \\
        w^0 & w^3 & w^6 & w^9   
           \end{array} \right]
    \left[ \begin{array}{c}
        x_0\\
        x_1\\
        x_2\\
        x_3
               \end{array}\right]
\label{dft}\end{equation}\\[10pt]
\vspace{.1in}
An example of the
second might be convolution where you are processing or filtering a
signal and staying in the same space or coordinate system.
\vspace{.1in}
\begin{equation}
    \left[ \begin{array}{c}
    y_0\\
    y_1\\
    y_2\\
    \vdots
    \end{array} \right]
    =
        \left[ \begin{array}{ccccc}
        h_0 & 0   & 0   & \cdots & 0\\
        h_1 & h_0 & 0   &  & \\
        h_2 & h_1 & h_0 &  & \\
        \vdots & & & & \vdots
        \end{array} \right]
            \left[ \begin{array}{c}
            x_0\\
            x_1\\
            x_2\\
            \vdots
        \end{array} \right].
\label{con}\end{equation}
\vspace{.1in}

A particularly powerful sequence of operations is to first change the basis
for a signal, then process the signal in this new basis, and finally return
to the original basis.  For example, the discrete Fourier transform (DFT)
of a signal is taken followed by setting some of the Fourier coefficients to
zero followed by taking the inverse DFT.

Another application of (\ref{1c}) is made in linear regression where the
input signals are rows of \A and the unknown weights of the hypothesis 
are in \x and the outputs are the elements of \bb.

\section{Change of Basis}
Consider the two views:
\ben
\item
The operation given in (\ref{1c}) can be viewed as \x
being a set of weights so that \bb is a
weighted sum of the columns of \A.  In other words, \bb will lie in the
space spanned by the columns of \A at a location determined by \x.  This
view is a composition of a signal from a set of weights as in (\ref{a1}) 
and (\ref{a3}) below.  If the vector ${\bf a_i}$ is the 
$i^{th}$ column of ${\bf A}$, it is illustrated by 
\begin{equation}
{\bf A x} = x_1 
    \left[ \begin{array}{c}
    \vdots\\
    {\bf a_1}\\
    \vdots
    \end{array} \right]
    + x_2 
    \left[ \begin{array}{c}
    \vdots\\
    {\bf a_2}\\
    \vdots
    \end{array} \right]
    + x_3 
    \left[ \begin{array}{c}
    \vdots\\
   {\bf  a_3}\\
    \vdots
        \end{array} \right]
   = {\bf b}.
\label{expan}\end{equation}
\item
An alternative view has \x being a signal vector
and with \bb being a vector whose entries are inner products of \x and the
rows of {\bf A}.   In other words, the elements of \bb are the projection
coefficients of \x onto the coordinates given by the rows of {\bf A}.  The
multiplication of a signal by this operator decomposes the signal and gives the
coefficients of the decomposition.  If ${\bf \bar{a}_j}$ is the $j^{th}$ row of ${\bf A}$
we have:
\begin{equation}
   b_1 = \left[\begin{array}{ccc} \dots {\bf \bar{a}_1} \dots \end{array}\right] 
             \left[\begin{array}{c} \vdots \\ {\bf x} \\ \vdots \end{array}\right]
\ \ \ \ \ \ \ \
   b_2 = \left[\begin{array}{ccc} \dots {\bf \bar{a}_2} \dots \end{array}\right] 
             \left[\begin{array}{c} \vdots \\ {\bf x} \\  \vdots \end{array}\right]
\ \ \ \ \ \ \ \  etc.
\label{ip}\end{equation}
Regression can be posed from this view with the input signal being the rows of {\bf A}.
\label{1d}\een

These two views of the operation as a decomposition of a signal or the
recomposition of the signal to or from a different basis system are
extremely valuable in signal analysis.  The ideas from linear algebra of subspaces, 
inner product, span, orthogonality, rank, etc. are all important here.  
The dimensions of the domain and
range of the operators may or may not be the same.  The matrices may or
may not be square and may or may not be of full rank \cite{halmos,strang0}.

\subsection{A Basis and Dual Basis}

A set of linearly independent vectors ${\bf x_n}$ forms a \emph{basis} for
a vector space if every vector \x in the space can be uniquely written
\be
    {\bf x} = \sum_n a_n \,{\bf x_n}
\label{a1}\ee
and the {\it dual basis} is defined as a set vectors ${\bf\tilde{x}_n}$ in that space allows a simple inner
product (denoted by parenthesis: $({\bf x,y})$) to calculate the expansion coefficients as
\be
    a_n =\, ( {\bf x, \tilde{x}_n} )\, = {\bf x^T \tilde{x}_n}
\label{a2}\ee

A basis expansion has enough vectors but none extra.  It is efficient in that no fewer expansion vectors
will represent all the vectors in the space but is fragil in that losing one coefficient or one basis vector
destroys the ability to exactly represent the signal by (\ref{a1}).
The expansion (\ref{a1}) can be written as a matrix operation
\be
    {\bf F \, a = x}
\label{a3}\ee
where the columns of {\bf F} are the basis vectors ${\bf x_n}$ and the vector {\bf a}
has the expansion coefficients $a_n$ as entries.  Equation (\ref{a2}) can
also be written as a matrix operation
\be
    {\bf \tilde{F} \, x = a}
\label{a4}\ee
which has the dual basis vectors as rows of ${\bf \tilde{F}}$.  From
(\ref{a3}) and (\ref{a4}), we have
\be
    {\bf F \tilde{F} \, x = x}
\label{a5}\ee
Since this is true for all ${\bf x}$,
\be
    {\bf F\, \tilde{F} = I}
\label{a6}\ee
or
\be
    {\bf \tilde{F} = F^{-1}}
\label{a61}\ee
which states the dual basis vectors are the rows of the inverse of the
matrix whose columns are the basis vectors (and vice versa).  When the
vector set is a basis, {\bf F} is necessarily square and from (\ref{a3})
and (\ref{a4}), one can show
\be
    {\bf F\, \tilde{F} = \tilde{F} \,F}.
\label{a7}\ee
Because this system requires two basis sets, the expansion basis and the
dual basis, it is called {\it biorthogonal}.

\subsection{Orthogonal Basis}

If the basis vectors are not only independent but orthonormal, the basis
set is its own dual and the inverse of {\bf F} is simply its transpose.
\be
    {\bf F^{-1} = \tilde{F} = F^T}
\label{a71}\ee
When done in Hilbert spaces, this decomposition is sometimes called an
abstract Fourier expansion \cite{halmos,halmos2,young3}.

\subsection{Parseval's Theorem}

Because many signals are digital representations of voltage, current, force, velocity, pressure, flow, etc.,
the inner product of the signal with itself (the norm squared) is a measure of the signal energy $q$.
\begin{equation}
			q = {\bf (x,x)} =  {\bf ||x||^2} = {\bf x^Tx} = \sum_{n=0}^{N-1} x_n^2
\label{2cc}\end{equation}
\vspace{10pt}
\emph{Parseval's theorem} states that if the basis system is orthogonal, then the norm squared (or ``energy'') 
is invarient across a change in basis.   If a change of basis is made with
\begin{equation}
			{\bf c = Ax}
\label{2cq}\end{equation}
then
\begin{equation}
	q = {\bf (x,x)} =  {\bf ||x||^2} = {\bf x^Tx} = \sum_{n=0}^{N-1} x_n^2 
	  = K {\bf (c,c)} =  K{\bf ||c||^2} = K{\bf c^Tc} = K\sum_{k=0}^{N-1} c_k^2
\label{2cf}\end{equation}
for some constant $K$ which can be made unity by normalization if desired.

For the discrete Fourier transform (DFT) of $x_n$ which is
\begin{equation}
			c_k = \frac{1}{N} \sum_{n=0}^{N-1} x_n e^{-j 2\pi nk/N}
\label{2cp}\end{equation}
the energy calculated in the time domain: $q = \sum_n x_n^2$ is equal to the norm squared
of the frequency coefficients: $q = \sum_k c_k^2$, within a multiplicative constant of $1/N$.  
This is because the basis functions of the Fourier transform are orthogonal:
``the sum of the squares is the square of the sum'' which means 
means the energy calculated in the time domain is the same as that calculated in the frequency domain. 
The energy of the signal (the square of the sum) is the sum of the energies at each frequency (the
sum of the squares).  Because of the orthogonal basis, the cross terms are zero.
Although one seldom directly uses Parseval's theorem, its truth is what make sense in talking about 
frequency domain filtering of a time domain signal.  A more general form is known as  the Plancherel theorem
\cite{frames}.

If a transformation is made on the signal with a non-orthogonal basis system, then Parseval's theorem does
not hold and the concept of energy does not move back and forth between domains.  We can get around some
of these restrictions by using frames rather than bases for expansions.

\subsection{Frames and Tight Frames}

In order to look at a more general expansion system than a basis and to generalize the ideas
of orthogonality and of energy
being calculated in the original expansion system or the transformed system, the concept of \emph{frame} is
defined.  A frame decomposition or representation is generally more robust and flexible than a
basis decomposition or representation but it requires more computation and memory \cite{heil3,waldron,frames}.
Sometimes a frame is called a redundant basis or representing an underdetermined or underspecified set of
equations.

If a set of vectors, ${\bf f_k}$, span a vector space (or subspace) but are not necessarily independent nor orthogonal, 
bounds on the energy in the transform can still be defined.  A set of vectors that span a vector space is called a 
\emph{frame} if two constants, $A$ and $B$  exist such that
\begin{equation}
		0 < A||{\bf x}||^2 \leq \sum_k |({\bf f_k,x})|^2 \leq B||{\bf x}||^2 < \infty
\label{2ce}\end{equation}
and the two constants are called the {\it frame bounds} for the system.  This can be written
\begin{equation}
		0 < A||{\bf x}||^2 \leq ||{\bf c}||^2 \leq B||{\bf x}||^2 < \infty
\label{2ce1}\end{equation}
where
\begin{equation}
		{\bf c= F x}
\label{2ce2}\end{equation}
If the ${\bf f_k}$ are linearly independent but not orthogonal, then the frame is a non-orthogonal basis.  If the 
${\bf f_k}$ are not independent
the frame is called redundant  since there are more than the minimum number of expansion vectors that
a basis would have.  If the
frame bounds are equal, $A=B$, the system is called a \emph{tight frame} and it has many of features of an orthogonal basis.  
If the bounds are equal to each other and to one, $A=B=1$, then the frame is a basis and is tight.  
It is, therefore, an orthogonal basis.

So a frame is a generalization of a basis and a tight frame is a generalization of an orthogonal basis.  If , $A = B$, the frame
is tight and we have a scaled Parseval's theorem:
\begin{equation}
		A||{\bf x}||^2 = \sum_k |({\bf f_k,x})|^2 
\label{2cf1}\end{equation}
If $A = B > 1$, then the number of expansion vectors are more than needed for a basis and $A$ is a measure of the redundancy
of the system (for normalized frame vectors).  For example, if there are three frame vectors in a two dimensional vector space, $A=3/2$.

A finite dimensional matrix version of the redundant case would have ${\bf F}$ in (\ref{a3}) with more columns
than rows but with full row rank.  For example
\begin{equation}
 \left[ \begin{array}{ccc}
        a_{00} & a_{01} & a_{02}\\
        a_{10} & a_{11} & a_{12}\\
        \end{array} \right]
            \left[ \begin{array}{c}
            x_0\\
            x_1\\
            x_2\\
                   \end{array} \right]
  =
    \left[ \begin{array}{c}
    b_0\\
    b_1\\
            \end{array} \right]
\label{3}\end{equation}
has three frame vectors as the columns of \A but in a two dimensional space.

The prototypical example is called the ``Mercedes-Benz'' tight frame where three frame 
vectors that are ${120^\circ}$ apart are used in a two-dimensional
plane and look like the Mercedes car hood ornament.  These three frame vectors must be as
far apart from each other as possible to be tight, hence the ${120^\circ}$ separation. 
But, they can be rotated any amount and remain tight \cite{waldron,jelena2} and, therefore,
are not unique.
\begin{equation}
 \left[ \begin{array}{ccc}
        1 & -0.5 & -0.5 \\
        0 & 0.866 & -0.866 \\
        \end{array} \right]
            \left[ \begin{array}{c}
            x_0\\
            x_1\\
            x_2\\
                   \end{array} \right]
  =
    \left[ \begin{array}{c}
    b_0\\
    b_1\\
            \end{array} \right]
\label{4}\end{equation}
In the next section, we will use the pseudo-inverse of ${\bf A}$ to find the optimal ${\bf x}$ for a given ${\bf b}$.

So the frame bounds $A$ and $B$ in (\ref{2ce}) are an indication of the redundancy of the expansion system $f_k$ and to
how close they are to being orthogonal or tight.  Indeed, (\ref{2ce}) is a sort of approximate Parseval's theorem 
\cite{young,id1,peiyeh,jelena,frames,waldron,ferreira,jelena3}.

The dual frame vectors are also not
unique but a set can be found such that (\ref{a4}) and, therefore,
(\ref{a5}) hold (but (\ref{a7}) does not).  A set of dual frame vectors
could be found by adding a set of arbitrary but independent rows to {\bf
F} until it is square, inverting it, then taking the first $N$ columns to
form ${\bf \tilde{F}}$ whose rows will be a set of dual frame vectors.
This method of construction shows the non-uniqueness of the dual frame
vectors.  This non-uniqueness is often resolved by optimizing some other
parameter of the system \cite{id1}.

If the matrix operations are implementing a frame decomposition and the
rows of {\bf F} are orthonormal, then ${\bf \tilde{F} = F^T}$ and the vector set
is  a  tight frame \cite{young,id1}.  If the frame vectors are
normalized to $||{\bf x_k}|| = 1$, the decomposition in (\ref{a1}) becomes
\be
   {\bf x} = \frac{1}{A} \sum_n
        ( {\bf x, \tilde{x}_n} ) \;{\bf x_n}
\label{a72}\ee
where the constant $A$ is a measure of the redundancy of the expansion
which has more expansion vectors than necessary \cite{id1}.

The matrix form is
\be
    {\bf x} = \frac{1}{A} \; {\bf F \, F^T \; x}
\label{a73}\ee
where ${\bf F}$ has more columns than rows.  Examples can be found in
\cite{bgwave}.

\subsection{Sinc Expansion as a Tight Frame}\index{sinc}

The Shannon sampling theorem \cite{csbcnxfd,yonina} can be viewied as an infinite dimensional signal expansion
where the \emph{sinc} functions are an orthogonal basis.  The sampling theorem 
with critical sampling, i.e. at the Nyquist rate, is the expansion:
\be
        g(t) \ = \  \sum_n g(Tn)\,
                \frac{\sin(\frac{\pi}{T}(t-Tn))}
                        {\frac{\pi}{T}(t-Tn)}
\label{a23r1}\ee
where the expansion coefficients are the samples and where 
the \emph{sinc} functions are easily shown to be orthogonal.

\emph{Over sampling} is 
an example of an infinite-dimensional tight frame \cite{marks,bgwave}.
If a function is over-sampled but the sinc functions remains consistent
with the upper spectral limit $W$, using $A$ as the amount of 
over-sampling, the sampling theorem becomes:
\be
        A W \ = \ \frac{\pi}{T}, \ \ \ \ \mbox{for $A \geq 1$}
\label{a23r2}\ee
and we have
\be
        g(t) \ = \ \frac{1}{A} \sum_n g(Tn)\,
                \frac{\sin(\frac{\pi}{AT}(t-Tn))}
                        {\frac{\pi}{AT}(t-Tn)}
\label{a23r3}\ee
where the sinc functions are no longer orthogonal.  In fact, they are
no longer a basis as they are not independent.  They are, however, a tight
frame and, therefore, have some of the characteristics of an orthogonal basis but with
a ``redundancy" factor $A$ as a multiplier in the formula \cite{bgwave} and
a generalized Parseval's theorem.
Here, moving from a basis to a frame (actually from an orthogonal basis
to a tight frame) is almost invisible. 

\subsection{Frequency Response of an FIR Digital Filter}

The discrete-time Fourier transform (DTFT) of the impulse response of an
FIR digital filter $h(n)$ is its frequency response.  The discrete Fourier transform (DFT) of 
$h(n)$ gives samples of the frequency response \cite{csbcnxfd}.  This is a powerful analysis tool
in digital signal processing (DSP)  and suggests that an inverse (or pseudoinverse) 
method could be useful for design \cite{csbcnxfd}.

\subsection{Conclusions}

Frames tend to be more robust than bases in tolerating errors and missing
terms.  They allow flexibility is designing wavelet systems \cite{id1} where
frame expansions are often chosen.

In an infinite dimensional vector space, if basis vectors are chosen such
that all expansions converge very rapidly, the basis is called an {\it
unconditional basis} and is near optimal for a wide class of signal
representation and processing problems.  This is discussed by Donoho in
\cite{dld3}.

Still another view of a matrix operator being a change of basis can be
developed using the eigenvectors  of an operator as
the basis vectors.  Then a signal can decomposed into its eigenvector
components which are then simply multiplied by the scalar eigenvalues to
accomplish the same task as a general matrix multiplication.  This is an
interesting idea but will not be developed here.

\section{Change of Signal}

If both \x and \bb in (\ref{1c}) are considered to be signals in the same coordinate or
basis system, the matrix operator \A is generally square.  It may or may
not be of full rank and it may or may not have a variety of other
properties, but both \x and \bb are viewed in the same coordinate system and
therefore are the same size.

One of the most ubiquitous of these is convolution where the input to a linear,
shift invariant system with impulse response $h(n)$ is calculated by (\ref{1c}) 
if ${\bf A}$ is the convolution matrix and ${\bf x}$ is the input \cite{csbcnxfd}.  
\begin{equation}
    \left[ \begin{array}{c}
    y_0\\
    y_1\\
    y_2\\
    \vdots
    \end{array} \right]
    =
        \left[ \begin{array}{ccccc}
        h_0 & 0   & 0   & \cdots & 0\\
        h_1 & h_0 & 0   &  & \\
        h_2 & h_1 & h_0 &  & \\
        \vdots & & & & \vdots
        \end{array} \right]
            \left[ \begin{array}{c}
            x_0\\
            x_1\\
            x_2\\
            \vdots
        \end{array} \right].
\label{con1}\end{equation}
It can also be calculated if ${\bf A}$ is the arrangement of the input and 
${\bf x}$ is the the impulse response.
\begin{equation}
    \left[ \begin{array}{c}
    y_0\\
    y_1\\
    y_2\\
    \vdots
    \end{array} \right]
    =
        \left[ \begin{array}{ccccc}
        x_0 & 0   & 0   & \cdots & 0\\
        x_1 & x_0 & 0   &  & \\
        x_2 & x_1 & x_0 &  & \\
        \vdots & & & & \vdots
        \end{array} \right]
            \left[ \begin{array}{c}
            h_0\\
            h_1\\
            h_2\\
            \vdots
        \end{array} \right].
\label{con2}\end{equation}
If the signal is periodic or if the DFT is being used, then what is called a
\emph{circulate} is used to represent cyclic convolution.  An example for
$N=4$ is the Toeplitz system
\begin{equation}
    \left[ \begin{array}{c}
    y_0\\
    y_1\\
    y_2\\
    y_3
    \end{array} \right]
    =
        \left[ \begin{array}{ccccc}
        h_0 & h_3 & h_2 & h_1 \\
        h_1 & h_0 & h_3 & h_2 \\
        h_2 & h_1 & h_0 & h_3 \\
        h_3 & h_2 & h_1 & h_0
        \end{array} \right]
            \left[ \begin{array}{c}
            x_0\\
            x_1\\
            x_2\\
            x_3
        \end{array} \right].
\label{con3}\end{equation}

One method of understanding and generating matrices of this sort is to
construct them as a product of first a decomposition operator, then a
modification operator in the new basis system, followed by a recomposition
operator.  For example, one could first multiply a signal by the DFT
operator which will change it into the frequency domain.  One (or more)
of the frequency coefficients could be removed (set to zero) and the
remainder multiplied by the inverse DFT operator to give a signal back in
the time domain but changed by having a frequency component removed.  That
is a form of signal filtering and one can talk about removing the energy of 
a signal at a certain frequency (or many) because of Parseval's theorem.

It would be instructive for the reader to make sense out of the cryptic
statement ``the DFT diagonalizes the cyclic convolution matrix" to add to
the ideas in this note.

\section{Factoring the Matrix ${\bf A}$}

For insight, algorithm development, and/or computational efficiency, it is sometime
worthwhile to factor ${\bf A}$ into a product of two or more matrices. For example,
the $DFT$ matrix \cite{csbcnxfft} illustrated in (\ref{dft}) can be factored into a product
of fairly sparce matrices.  If fact, the fast Fourier transform (FFT) can be derived by
factoring the DFT matrix into $N \log(N)$ factors (if $N=2^m$), each requiring order $N$ multiplies. 
This is done in \cite{csbcnxfft}.

Using eigenvalue theory \cite{strang0}, a full rank square matrix can be factored into a product 
\begin{equation}
		{\bf A V = V \Lambda}
\label{ev}\end{equation}
where ${\bf V}$ is a matrix with columns of the eigenvectors of ${\bf A}$ and 
${\bf \Lambda}$ is a diagonal matrix with the eigenvalues along the diagonal. The
inverse is a method to ``diagonalize" a matrix
\begin{equation}
		{\bf \Lambda = V^{-1} A V}
\label{ev1}\end{equation}
If a matrix has ``repeated eigenvalues", in other words, two or more of the $N$
eigenvalues have the same value but less than $N$ indepentant eigenvectors, it is 
not possible to diagonalize the matrix but
an ``almost" diagonal form called the {\it Jordan normal form} can be acheived.
Those details can be found in most books on matrix theory \cite{strang}.

A more general decompostion is the singular value decomposition (SVD) which is
similar to the eigenvalue problem but allows rectangular matrices.
It is particularly valuable for expressing the pseudoinverse in a simple form and
in making numerical calculations \cite{Trefethen2}.

\section{State Equations}
If our matrix multiplication equation is a vector differential equation (DE) of the form
\be
    {\bf \dot{x} = A x}
\label{1e}\ee
or for difference equations and discrete-time signals or digital signals,
\be
    {\bf x}(n+1) = {\bf A x}(n)
\label{1f}\ee
an inverse or even pseudoinverse will not solve for \x.  A different
approach must be taken \cite{dorf} and different properties and tools from
linear algebra will be used. The solution of this first order vector DE is a 
coupled set of solutions of first order DEs.  If a change of
basis is made so that ${\bf A}$ is diagonal (or Jordan form), equation (\ref{1e}) becomes a
set on uncoupled (or almost uncoupled in the Jordan form case) first order DEs and we know the solution of a first order
DE is an exponential.  This requires consideration of the eigenvalue problem,
diagonalization, and solution of scalar first order DEs \cite{dorf}.

State equations are often used to model or describe a system such as a 
control system or a digital filter or a numerical algorithm \cite{dorf,zadeh2}.

\chapter{Solutions of Simultaneous Linear Equations}\label{approx}
The second problem posed in the introduction is basically the solution of
simultaneous linear equations \cite{ls,albert,israel} which is fundamental to 
linear algebra \cite{hefferon,strang0,moler5} and
very important in diverse areas of applications in mathematics, numerical
analysis, physical and social sciences, engineering, and business.  Since
a system of linear equations may be over or under determined in a variety
of ways, or may be consistent but ill conditioned, a comprehensive theory
turns out to be more complicated than it first appears.  Indeed, there is
a considerable literature on the subject of {\it generalized inverses} or
{\it pseudo-inverses}.  The careful statement and formulation of the general
problem seems to have started with Moore \cite{ehmoore} and Penrose
\cite{penrose1,penrose2} and developed by many others.  
Because the generalized solution of simultaneous equations
is often defined in terms of minimization of an equation error, the techniques
are useful in a wide variety of approximation and 
optimization problems \cite{bjorck,dgl1} as well as signal processing.

The ideas are presented here in terms of finite dimensions using
matrices.  Many of the ideas extend to infinite dimensions using Banach
and Hilbert spaces \cite{riesz,oden,young} in functional analysis.

\section{The Problem}\label{problem}

Given an $M$ by $N$ real matrix ${\bf A}$ and an $M$ by $1$ vector ${\bf
b}$, find the $N$ by $1$ vector ${\bf x}$ when
\begin{equation}
 \left[ \begin{array}{ccccc}
        a_{11} & a_{12} & a_{13} & \cdots & a_{1N}\\
        a_{21} & a_{22} & a_{23} &  & \\
        a_{31} & a_{32} & a_{33} &  & \\
        \vdots & & & & \vdots \\
        a_{M1} &  &  & \cdots & a_{MN}
        \end{array} \right]
            \left[ \begin{array}{c}
            x_1\\
            x_2\\
            x_3\\
            \vdots\\
            x_{N}\\
                  \end{array} \right]
  =
    \left[ \begin{array}{c}
    b_1\\
    b_2\\
    b_3\\
    \vdots\\
    b_{M}\\
         \end{array} \right]
\label{0}\end{equation}
or, using matrix notation,
\be
    {\bf A x = b}
\label{1a}\ee
If ${\bf b}$ does not lie in the range space of ${\bf A}$ (the space
spanned by the columns of ${\bf A}$), there is no exact solution to
(\ref{1a}), therefore, an approximation problem can be posed by 
minimizing an equation error defined by
\be
    {\bf \varepsilon = A x - b}.
\label{2a}\ee
A generalized solution (or an optimal approximate solution) to (\ref{1a}) is usually considered to be an ${\bf x}$
that minimizes some norm of $\varepsilon$. If that problem does not have a unique solution, further 
conditions, such as also minimizing the norm of ${\bf x}$, are imposed. The $l_2$ or root-mean-squared 
error or Euclidean norm is
 $\sqrt{\bf \varepsilon^{T*} \varepsilon}$ and minimization sometimes has an analytical solution.  
Minimization of other norms such as ${l_{\infty}}$ (Chebyshev) or ${l_1}$ require iterative solutions. 
The general $l_p$ norm is defined as $q$ where
\be
	q = ||x||_p = \biggl( {\sum_n |x(n)|^p \biggr)^{1/p}}
\label{3q}\ee
for $1 < p < \infty$ and a ``pseudonorm" (not convex) for $0 < p < 1$.
These can sometimes be evaluated using IRLS (iterative reweighted least squares) algorithms  \cite{bb1,bbs,irlsvb,gorodnitsky,DaubSparse}.

If there is a non-zero solution of the homogeneous equation
\be
    {\bf A x = 0},
\label{3a}\ee
then (\ref{1a}) has infinitely many generalized solutions in the sense that any particular
solution of (\ref{1a}) plus an arbitrary scalar times any non-zero solution of
(\ref{3a}) will have the same error in (\ref{2a}) and, therefore, is also a
generalized solution.  The number of families of solutions is the dimension 
of the null space of ${\bf A}$.

This is analogous to the classical solution of linear, constant coefficient differential equations
where the total solution consists of a particular solution plus arbitrary constants times the
solutions to the homogeneous equation.  The constants are determined from the initial 
(or other) conditions of the solution to the differential equation.

\section{Ten Cases to Consider}
Examination of the basic problem shows there are ten cases \cite{ls} listed in 
Figure 1 to be considered.  
These depend on the shape of the $M$ by $N$ real matrix ${\bf A}$,
the rank $r$ of ${\bf A}$, and whether \bb is in the span of the columns of ${\bf A}$.
\vspace{.2in}
\begin{itemize}
\item 1a. $M = N = r$: One solution with no error, ${\varepsilon}$.
\item 1b. $M = N > r$: ${\bf b} \: \in span\{{\bf A}\}$: Many solutions with ${\bf \varepsilon  = 0}$.
\item 1c. $M = N > r$: ${\bf b} \: not \in span\{{\bf A}\}$: Many solutions with the same minimum error.
\item 2a. $M > N = r$: ${\bf b} \: \in span\{{\bf A}\}$: One solution ${\bf \varepsilon  = 0}$.
\item 2b. $M > N = r$: ${\bf b} \: not \in span\{{\bf A}\}$: One solution with minimum error.
\item 2c. $M > N > r$: ${\bf b} \: \in span\{{\bf A}\}$: Many solutions with ${\bf \varepsilon  = 0}$.
\item 2d. $M > N > r$: ${\bf b} \: not \in span\{{\bf A}\}$: Many solutions with the same minimum error.
\item 3a. $N > M = r$: Many solutions with ${\bf \varepsilon  = 0}$.
\item 3b. $N > M > r$: ${\bf b} \: \in span\{{\bf A}\}$: Many solutions with ${\bf \varepsilon  = 0}$
\item 3c. $N > M > r$: ${\bf b} \: not \in span\{{\bf A}\}$: Many solutions with the same minimum error.
\end{itemize}
{\bf Figure 1.} Ten Cases for the Pseudoinverse.\\[5pt]
Here we have: 
\begin{itemize}
\item case 1 has the same number of equations as unknowns ({\bf A} is square, $M=N$),
\item case 2 has more equations than unknowns, therefore, is over specified  ({\bf A} is taller than wide, $M>N$),
\item case 3 has fewer equations than unknowns, therefore, is underspecified  ({\bf A} is wider than tall $N>M$).
\end{itemize}
This is a setting for frames and sparse representations.

In case 1a and 3a, ${\bf b}$ is necessarily in the span of ${\bf A}$. 
In addition to these classifications, the possible orthogonality of the
columns or rows of the matrices gives special characteristics.

\section{Examples}

Case 1: Here we see a 3 x 3 square matrix which is an example of case 1 in Figure 1 and 2.
\begin{equation}
 \left[ \begin{array}{ccc}
        a_{11} & a_{12} & a_{13} \\
        a_{21} & a_{22} & a_{23} \\
        a_{31} & a_{32} & a_{33} 
        \end{array} \right]
            \left[ \begin{array}{c}
            x_1\\
            x_2\\
            x_3
                  \end{array} \right]
  =
    \left[ \begin{array}{c}
    b_1\\
    b_2\\
    b_3
         \end{array} \right]
\label{1aa}\end{equation} 
If the matrix has rank 3, then the \bb vector will necessarily be in the space spanned
by the columns of \A which puts it in case 1a.  This can be solved for \x by inverting
\A or using some more robust method. If the matrix has rank 1 or 2, the \bb
may or may not lie in the spanned subspace, so the classification will be 1b or 1c and
minimization of $||x||_2^2$ yields a unique solution.

Case 2: If \A is 4 x 3, then we have more equations than unknowns or the overspecified 
or overdetermined case.
\begin{equation}
 \left[ \begin{array}{ccc}
        a_{11} & a_{12} & a_{13} \\
        a_{21} & a_{22} & a_{23} \\
        a_{31} & a_{32} & a_{33} \\
	 a_{41} & a_{42} & a_{43}
        \end{array} \right]
            \left[ \begin{array}{c}
            x_1\\
            x_2\\
            x_3
                  \end{array} \right]
  =
    \left[ \begin{array}{c}
    b_1\\
    b_2\\
    b_3\\
    b_4
         \end{array} \right]
\label{1aac}\end{equation} 
If this matrix has the maximum rank of 3, then we have case 2a or 2b depending on
whether \bb is in the span of \A or not.  In either case, a unique solution \x exists
which can be found by (\ref{6a}) or (\ref{pi2}).  For case 2a, we have a single exact solution with
no equation error, ${\bf \epsilon =0}$ just as case 1a.  For case 2b, we have a single
optimal approximate solution with the least possible equation error.  If the matrix has
rank 1 or 2, the classification will be 2c or 2d and minimization of $||x||_2^2$ yelds a unique solution.

Case 3: If \A is 3 x 4, then we have more unknowns than equations or the underspecified case.
\begin{equation}
 \left[ \begin{array}{cccc}
        a_{11} & a_{12} & a_{13} & a_{14} \\
        a_{21} & a_{22} & a_{23} & a_{24} \\
        a_{31} & a_{32} & a_{33} & a_{34} 
        \end{array} \right]
            \left[ \begin{array}{c}
            x_1\\
            x_2\\
            x_3\\
	     x_4
                  \end{array} \right]
  =
    \left[ \begin{array}{c}
    b_1\\
    b_2\\
    b_3
         \end{array} \right]
\label{1aad}\end{equation} 
If this matrix has the maximum rank of 3, then we have case 3a and \bb must be in 
the span of \A.  For this case, many exact solutions \x exist, all having zero equation 
error and a single one can be found with minimum solution norm $||{\bf x}||$
using (\ref{6b}) or (\ref{pi3}). If the matrix has rank 1 or 2, the classification will be 3b or 3c.

\section{Solutions}

There are several assumptions or side conditions that could be used in
order to define a useful unique solution of (\ref{1a}).  The side conditions used
to define the Moore-Penrose pseudo-inverse are that the $l_2$ norm squared of the equation error
${\varepsilon}$ be minimized and, if there is ambiguity (several solutions
with the same minimum error), the $l_2$ norm squared of ${\bf x}$ also be minimized.  A
useful alternative to minimizing the norm of ${\bf x}$ is to require
certain entries in ${\bf x}$ to be zero (sparse) or fixed to some non-zero value
(equality constraints).

In using sparsity in posing a signal processing problem (e.g. compressive sensing), an
$l_1$ norm can be used (or even an $l_0$ ``pseudo norm'') to obtain solutions with 
zero components if possible \cite{donoho6,IvanSparse}.

In addition to using side conditions to achieve a unique solution, side
conditions are sometimes part of the original problem.  One interesting case
requires that certain of the equations be satisfied with no error and the
approximation be achieved with the remaining equations.

\section{Moore-Penrose Pseudo-Inverse}

If the $l_2$ norm is used, a unique generalized solution to (\ref{1a}) always exists such that the
norm squared of the equation error ${\bf \varepsilon^{T*} \varepsilon}$ and the norm squared of the
solution ${\bf x^{T*}x}$ are both minimized.  This solution is denoted by
\be
    {\bf x = A^+b}
\label{4a}\ee
where ${\bf A^+}$ is called the Moore-Penrose inverse \cite{albert} of ${\bf A}$ (and is also called
the generalized inverse \cite{israel} and the pseudoinverse \cite{albert})

Roger Penrose \cite{penrose2} showed that for all ${\bf A}$,
there exists a unique ${\bf A^+}$ satisfying the four conditions:

$${\bf A A^+ A = A}$$
\begin{equation}{\bf A^+ A A^+ = A^+}\label{prc}\end{equation}
$${\bf [A A^+]^* = A A^+}$$
$${\bf [A^+ A]^* = A^+ A}$$

There is a large literature on this problem.  Five useful books are
\cite{ls,albert,israel,campbell,raomitra}.  The Moore-Penrose
pseudo-inverse can be calculated in Matlab \cite{ml} by the
{\tt pinv(A,tol)} function which uses a singular value decomposition 
(SVD) to calculate it.  There are a variety of other numerical methods
given in the above references where each has some advantages and some
disadvantages.

\section{Properties}

For cases 2a and 2b in Figure 1, the following $N$ by $N$ system of equations
called the {\it normal equations} \cite{albert,ls} have a unique minimum squared 
equation error solution (minimum $\epsilon^T\epsilon$).  Here we have the
over specified case with more equations than unknowns.
A derivation is outlined in Section \ref{der}, equation (\ref{gradzero2}) below.
\be
    {\bf A^{T*}A x = A^{T*} b}
\label{5a}\ee
The solution to this equation is often used in least squares approximation
problems.  For these two cases ${\bf A}^T{\bf A}$ is non-singular and the $N$ by $M$ pseudo-inverse is simply,
\be
    {\bf A^+ = [A^{T*}A]^{-1} A^{T*}}.
\label{6a}\ee
A more general problem can be solved by minimizing the weighted equation error, 
${\bf \epsilon^T W^T W \epsilon}$ where ${\bf W}$ is a positive semi-definite
diagonal matrix of the error weights.  The solution to that problem \cite{israel} is
\be
    {\bf A^+ = [A^{T*} W^{T*} W A]^{-1} A^{T*} W^{T*} W}.
\label{6aw}\ee
For the case 3a in Figure 1 with more unknowns than equations, ${\bf A}{\bf A}^T$ is non-singular and  
has a unique minimum norm solution, ${\bf ||x||}$. The $N$ by $M$ pseudoinverse is simply,
\be
    {\bf A^+ = A^{T*}[A A^{T*}]^{-1} }.
\label{6b}\ee
with the formula for the minimum weighted solution norm $||x||$ is
\be
    {\bf A^+  = [W^T W]^{-1} A^T \biggl[ A [W^T W]^{-1}A^T\biggr]^{-1} }.
\label{6bw}\ee
For these three cases, either (\ref{6a}) or (\ref{6b}) can be directly calculated, but
not both.  However, they are equal so you simply use the one with the non-singular
matrix to be inverted.  The equality can be shown from
an equivalent definition \cite{albert} of the pseudo-inverse given
in terms of a limit by
\be
   {\bf A^+} = \lim_{\delta \rightarrow 0} {\bf [A^{T*} A + \delta^2 I]^{-1} A^{T*}}
             = \lim_{\delta \rightarrow 0} {\bf A^{T*} [A A^{T*} + \delta^2 I]^{-1}}.
\label{7a}\ee
For the other 6 cases, SVD or other approaches must be used. 
Some properties \cite{albert,campbell} are:
\begin{itemize}
\item ${\bf [A^+]^+ = A}$
\item ${\bf [A^+]^* = [A^*]^+}$
\item ${\bf [A^* A]^+ = A^+ A^{*+}}$
\item $\lambda^+ = 1/\lambda$ for $\lambda \neq 0$ else $\lambda^+ = 0$
\item ${\bf A^+ = [A^* A]^+ A^* = A^* [A A^*]^+}$
\item ${\bf A^* = A^* A A^+ = A^+ A A^*}$
\end{itemize}
It is informative to consider the range and null spaces \cite{campbell}
 of ${\bf A}$ and ${\bf A^+}$
\begin{itemize}
\item $R({\bf A}) = R({\bf A} {\bf A}^+) = R({\bf A} {\bf A}^*)$
\item $R({\bf A}^+) = R({\bf A}^*) = R({\bf A}^+ {\bf A}) = R({\bf A}^* {\bf A})$
\item $R(I - {\bf A} {\bf A}^+) = N({\bf A} {\bf A}^+) = N({\bf A}^*) = N({\bf A}^+) = R({\bf A})^{\perp}$
\item $R(I - {\bf A}^+ {\bf A}) = N({\bf A}^+ {\bf A}) = N({\bf A}) = R({\bf A}^*)^{\perp}$
\end{itemize}

\section{The Cases with Analytical Soluctions}

The four Penrose equations in (\ref{prc}) are remarkable in defining a unique 
pseudoinverse for any {\bf A} with any shape, any rank, for any of
the ten cases listed in Figure 1.  
However, only four cases of the ten have analytical solutions (actually, all do if you use SVD).
\begin{itemize}
\item If ${\bf A}$ is case 1a, (square and nonsingular), then \be{\bf A^+ = A^{-1}}\label{pi1}\ee 
\item If ${\bf A}$ is case 2a or 2b, (over specified) then \be{\bf A^+ = [A^T A]^{-1} A^T }\label{pi2}\ee
\item If ${\bf A}$ is case 3a, (under specified) then  \be{\bf A^+ =  A^T [ A A^T]^{-1}}\label{pi3}\ee
\end{itemize}
{\bf Figure 2.} Four Cases with Analytical Solutions\\[5pt]

Fortunately, most practical cases are one of these four but even then, it is generally
faster and less error prone to use special techniques on the normal equations 
rather than directly calculating the inverse matrix.  Note the matrices
to be inverted above are all $r$ by $r$ ($r$ is the rank) and nonsingular.  In the other
six cases from the ten in Figure 1, these would be singular, so alternate methods such
as SVD must be used \cite{ls,albert,israel}.

In addition to these four cases with ``analytical'' solutions, we can pose a more
general problem by asking for an optimal approximation with a weighted norm \cite{israel} 
to emphasize or de-emphasize certain components or range of equations.
\begin{itemize}
\item If ${\bf A}$ is case 2a or 2b, (over specified) then the weighted error pseudoinverse is
        \be{\bf A^+ = [A^{T*} W^{T*} W A]^{-1} A^{T*} W^{T*}W}\label{pi2w}\ee
\item If ${\bf A}$ is case 3a, (under specified) then the weighted norm pseudoinverse is
        \be{\bf A^+  = [W^T W]^{-1} A^T \biggl[ A [W^T W]^{-1}A^T\biggr]^{-1}}\label{pi3w}\ee
\end{itemize}
{\bf Figure 3.} Three Cases with Analytical Solutions and Weights\\[5pt]

These solutions to the weighted approxomation problem are useful in their own right
but also serve as the foundation to the Iterative Reweighted Least Squares (IRLS)
algorithm developed in the next chapter.

\section{Geometric interpretation and Least Squares Approximation}

A particularly useful application of the pseudo-inverse of a matrix is to
various least squared error approximations \cite{ls,bjorck}.  A geometric view of the
derivation of the normal equations can be helpful.  If ${\bf b}$ does not
lie in the range space of ${\bf A}$, an error vector is defined as the
difference between ${\bf A x}$ and ${\bf b}$.  A geometric picture of
this vector makes it clear that for the length of $\varepsilon$ to be
minimum, it must be orthogonal to the space spanned by the columns of
${\bf A}$.  This means that ${\bf A^* \varepsilon = 0}$.  If both sides of
(\ref{1a}) are multiplied by ${\bf A^*}$, it is easy to see that the normal
equations of (\ref{5a}) result in the error being orthogonal to the columns of
${\bf A}$ and, therefore its being minimal length.  If ${\bf b}$ does lie
in the range space of ${\bf A}$, the solution of the normal equations
gives the exact solution of (\ref{1a}) with no error.

For cases 1b, 1c, 2c, 2d, 3a, 3b, and 3c, the homogeneous equation
(\ref{3a}) has non-zero solutions.  Any vector in the space spanned by
these solutions (the null space of ${\bf A}$) does not contribute to the equation
error $\varepsilon$ defined in (\ref{2a}) and, therefore, can be added to
any particular generalized solution of (\ref{1a}) to give a family of
solutions with the same approximation error.  If the dimension of the null
space of ${\bf A}$ is $d$, it is possible to find a unique generalized
solution of (\ref{1a}) with $d$ zero elements.  The non-unique solution
for these seven cases can be written in the form \cite{israel}.
\be
    {\bf x = A^+b + [I - A^+ A] y}
\label{8a}\ee
where ${\bf y}$ is an arbitrary vector.  The first term is the minimum
norm solution given by the Moore-Penrose pseudo-inverse ${\bf A^+}$ and
the second is a contribution in the null space of ${\bf A}$.  For the 
minimum $||x||$, the vector ${{\bf y} = 0}$.

\subsection{Derivations}\label{der}

To derive the necessary conditions for minimizing $q$ in the overspecified case, we
differentiate $q = {\bf \epsilon^T \epsilon}$ with respect to ${\bf x}$ and set that
to zero.  Starting with the error
\begin{equation}
		q = {\bf \epsilon^T \epsilon} = {\bf [Ax-b]^T [Ax-b]}
		= {\bf x^T A^T A x  - x^T A^T b - b^T A x + b^T b}
\label{gradzero1}\end{equation}
\begin{equation}
		q = {\bf x^T A^T A x  - 2 x^T A^T b + b^T b}
\label{gradzero1a}\end{equation}
and taking the gradient or derivative gives 
\begin{equation}
		\nabla_{\bf x} q = 2 {\bf A^T A x} - 2 {\bf A^T b } = {\bf 0}
\label{gradzero2}\end{equation}
which are the normal equations in (\ref{5a}) and the pseudoinverse in (\ref{6a}) and (\ref{pi2}).

If we start with the weighted error problem 
\begin{equation}
		q = {\bf \epsilon^T W^T W \epsilon} = {\bf [Ax-b]^T W^T W  [Ax-b]}
\label{gradzero3}\end{equation}
using the same steps as before gives the normal equations for the minimum weighted squared error as
\begin{equation}
		{\bf A^TW^T W A x} = {\bf A^T W^T W b } 
\label{gradzero4}\end{equation}
and the pseudoinverse as 
\begin{equation}
		{\bf  x = [A^TW^T W A]^{-1} A^T W^T W b } 
\label{gradzero5}\end{equation}

To derive the necessary conditions for minimizing the Euclidian norm $||x||_2$ when
there are few equations and many solutions to (\ref{0}), we define a Lagrangian
\begin{equation}
	\mathcal{L}({\bf x,\mu}) = ||{\bf W x}||_2^2 + {\bf \mu^T(Ax-b)}
\label{lag}\end{equation}
take the derivatives in respect to both \x and ${\bf \mu}$ and set them to zero.
\begin{equation}
	{\bf \nabla_ x \mathcal{L} = 2 W^T W x + A^T \mu = 0}
\label{lag1}\end{equation}
and
\begin{equation}
	{\bf \nabla_\mu \mathcal{L} = Ax-b = 0}
\label{lag2}\end{equation}
Solve these two equation simultaneously for ${\bf x}$ eliminating ${\bf \mu}$ gives
the pseudoinverse in (\ref{6b}) and (\ref{pi3}) result.
\begin{equation}
		{\bf  x = [W^T W]^{-1} A^T \biggl[ A [W^T W]^{-1}A^T\biggr]^{-1} b } 
\label{lag3}\end{equation}

Because the weighting matrices ${\bf W}$ are diagonal and real, multiplication and 
inversion is simple.  These equations are used in the Iteratively Reweighted Least
Squares (IRLS) algorithm described in another section.

\section{Regularization}

To deal with measurement error and data noise, a process called ``regularization" is
sometimes used \cite{golub,bjorck,neumaier,hansen}.

\section{Least Squares Approximation with Constraints}

The solution of the overdetermined simultaneous equations is generally a
least squared error approximation problem.  A particularly interesting and
useful variation on this problem adds inequality and/or equality
constraints.  This formulation has proven very powerful in solving the
constrained least squares approximation part of FIR filter design
\cite{slb2}.  The equality constraints can be taken into account by using
Lagrange multipliers and the inequality constraints can use the
Kuhn-Tucker conditions \cite{fletcher,strang0,dgl2}. The iterative
reweighted least squares (IRLS) algorithm described in the next chapter
can be modified to give results which are an optimal constrained least p-power
solution \cite{bbs2,csb20,bbs}.

\section{Conclusions}
There is remarkable structure and subtlety in the apparently simple
problem of solving simultaneous equations and considerable insight can be
gained from these finite dimensional problems.  These notes have
emphasized the $l_2$ norm but some other such as $l_\infty$ and
$l_1$ are also interesting.   The use of sparsity \cite{IvanSparse} is particularly interesting
as applied in Compressive Sensing \cite{csrichb,donoho5} and in the sparse FFT \cite{hassanieh}. 
There are also interesting and important applications in
infinite dimensions.  One of particular interest is in signal analysis
using wavelet basis functions \cite{id1}.  The use of weighted error and
weighted norm pseudoinverses provide a base for iterative reweighted least
squares (IRLS) algorithms.

\chapter{Approximation with Other Norms and Error Measures}

Most of the discussion about the approximate solutions to ${\bf Ax=b}$ are about
 the result of minimizing the $l_2$ equation error $||Ax-b||_2$ 
and/or the $l_2$ norm of the solution $||{\bf x}||_2$ because in some cases that can be done 
by analytic formulas and also because the $l_2$ norm has a energy interpretation.
However, both the $l_1$ and the $l_\infty$ \cite{cheney} have well known applications that are
important \cite{DaubSparse,csbcnxfd} and the more general $l_p$ error 
is remarkably flexible \cite{bb1,bbs}.  Donoho has shown \cite{donoho71} that
$l_1$ optimization gives essentially the same sparsity as the true sparsity measure in $l_0$.

In some cases, one uses a different  norm for the minimization of
the equation error than the one for minimization of the solution norm.  And in
other cases, one minimizes a weighted error to emphasize some
equations relative to others \cite{israel}.  A modification allows minimizing according
to one norm for one set of equations and another for a different set.  A more general
error measure than a norm can be used which used a polynomial error \cite{bbs} which
does not satisfy the scaling requirement of a norm, but is convex.  One could even use the
so-called $l_p$ norm for ${1 > p > 0}$ which is not even convex but is an interesting
tool for obtaining sparse solutions. 

\begin{figure}[h]
    \psfig{figure=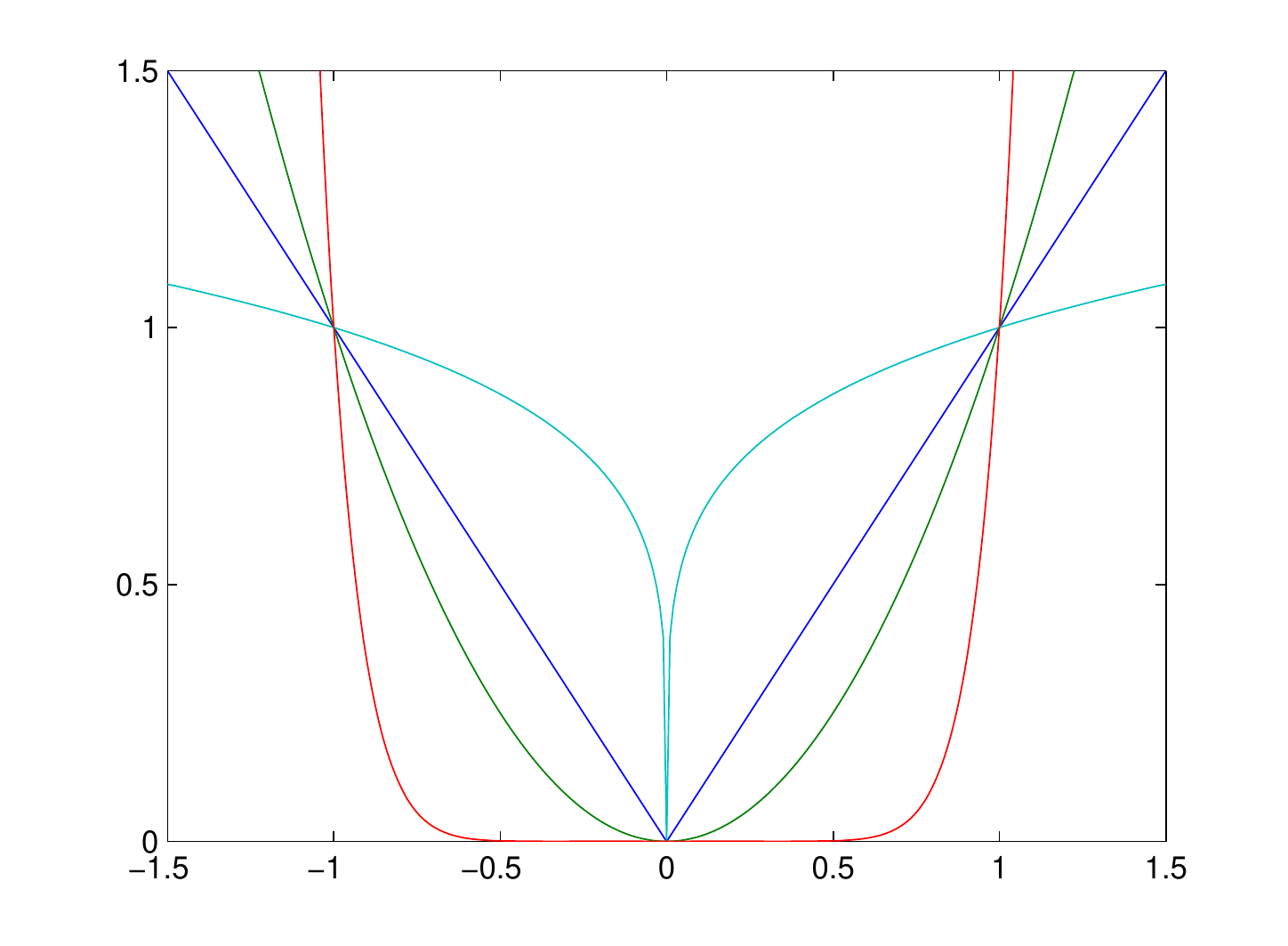,width=5.0in}
    \caption{Different $l_p$ norms: p = .2, 1, 2, 10.}
    \label{norm}
\end{figure}

Note from the figure how the $l_{10}$ norm puts a large penalty on large errors.
This gives a Chebyshev-like solution. The $l_{0.2}$ norm puts a large penalty on
small errors making them tend to zero.  This (and the $l_1$ norm) give a sparse solution.

\section{The $L_p$ Norm Approximation}

The {\bf IRLS} (iterative reweighted least squares) algorithm allows an iterative
algorithm to be built from the analytical solutions of the weighted least squares 
with an iterative reweighting to converge to the optimal $l_p$ approximation \cite{bjorck}.

\subsection{The Overdetermined System with more Equations than Unknowns}

If one poses the $l_p$ approximation problem in solving an overdetermined set of
equations (case 2 from Chapter 3), it comes from defining the equation error vector 
\begin{equation}
		\bf{e} = \bf{A x}-\bf{b}
\label{pn0}\end{equation}
and minimizing the p-norm

\begin{equation}
		||{\bf e}||_p = \biggl( \sum_n |e_n|^p\biggr)^{1/p}
\label{pn}\end{equation}
or
\begin{equation}
		||\bf{e}||_p^p = \sum_n |e_n|^p
\label{pn1}\end{equation}
neither of which can we minimize easily.  However, we do have formulas \cite{israel} to find the 
minimum of the weighted squared error
\begin{equation}
		||\bf{W e}||_2^2 =  \sum_n w_n^2 |e_n|^2
\label{pn2}\end{equation}
one of which is derived in Section \ref{der}, equation (\ref{gradzero2}) and is
\begin{equation}
		{\bf x} = [{\bf A^T W^T W A}]^{-1} {\bf A^T W^T W b}
\label{pn3}\end{equation}
where ${\bf W}$ is a diagonal matrix of the error weights, $w_n$.  From
this, we propose the iterative reweighted least squared (IRLS) error algorithm
which starts with unity weighting, ${\bf W}= {\bf I}$, solves for an
initial ${\bf x}$ with (\ref{pn3}), calculates a new error from(\ref{pn0}),
which is then used to set a new weighting matrix ${\bf W}$ 
\begin{equation}
		{\bf W} = diag(w_n)^{(p-2)/2}
\label{pn4}\end{equation}
to be used in the next iteration of (\ref{pn3}).
Using this, we find a new solution ${\bf x}$ and repeat until convergence
(if it happens!).

This core idea has been repeatedly proposed and developed in different
application areas over the past 50 years with a variety of success \cite{bjorck}.  Used
in this basic form, it reliably converges for ${2 < p < 3}$.  In 1990, a
modification was made to partially update the solution each iteration with
\begin{equation}
		{\bf x}(k) = q{\bf \hat{x}}(k) + (1-q) {\bf x}(k-1)
\label{pn5}\end{equation}
where ${\bf \hat{x}}$ is the new weighted least squares solution of (\ref{gradzero5})
which is used to partially update the previous value ${\bf x}(k-1)$ 
using a convergence up-date factor ${0 < q < 1}$ which gave convergence
over a larger range of around ${1.5 < p < 5}$ but but it was slower.  

A second improvement showed that a specific up-date factor of
\begin{equation}
		q = \frac{1}{p-1}
\label{pn6}\end{equation}
significantly increased the speed of convergence.  With this particular factor, the
algorithm becomes a form of Newton's method which has quadratic convergence.

A third modification applied homotopy \cite{bbs2,stonick3,stonick4,sieradski} 
by starting with a value for $p$ which is equal to $2$ and increasing it each iteration 
(or each few iterations) until it reached the desired value, or, in the case of $p<2$, decrease it.
This made a significant increase in both the
range of $p$ that allowed convergence and in the speed of calculations.  Some
of the history and details can be found applied to digital filter design in \cite{bb1,bbs}.

A Matlab program that implements these ideas applied to our pseudoinverse
problem with more equations than unknowns (case 2a) is:

\begin{verbatim}
% m-file IRLS1.m to find the optimal solution to Ax=b 
%  minimizing the L_p norm ||Ax-b||_p, using IRLS.
%  Newton iterative update of solution, x, for  M > N.
%  For 2<p<infty, use homotopy parameter K = 1.01 to 2  
%  For 0<p<2, use K = approx 0.7 - 0.9
%  csb 10/20/2012
function x = IRLS1(A,b,p,K,KK)
if nargin < 5, KK=10;  end; 
if nargin < 4, K = 2;  end;
if nargin < 3, p = 10; end;
pk = 2;                                      % Initial homotopy value
x  = pinv(A)*b;                              % Initial L_2 solution
E = [];
for k = 1:KK                                 % Iterate
   if p >= 2, pk = min([p, K*pk]);           % Homotopy change of p
      else pk = max([p, K*pk]); end
   e  = A*x - b;                             % Error vector
   w  = abs(e).^((pk-2)/2);                  % Error weights for IRLS
   W  = diag(w/sum(w));                      % Normalize weight matrix
   WA = W*A;                                 % apply weights 
   x1  = (WA'*WA)\(WA'*W)*b;                 % weighted L_2 sol.
   q  = 1/(pk-1);                            % Newton's parameter
   if p > 2, x = q*x1 + (1-q)*x; nn=p;       % partial update for p>2
      else x = x1; nn=2; end                 % no partial update for p<2
   ee = norm(e,nn);   E = [E ee];            % Error at each iteration
end
plot(E)
\end{verbatim}

This can be modified to use different $p$'s in different bands of equations or to 
use weighting only when the error exceeds a certain threshold to achieve a
constrained LS approximation \cite{bb1,bbs,irlsvb}.  Our work was originally
done in the context of filter design but others have done similar things in
sparsity analysis \cite{gorodnitsky,DaubSparse,yagle11}. 

This is presented as applied to the overdetermined system (Case 2a and 2b) but can also
be applied to other cases.  A particularly important application of this section is to
the design of digital filters.

\subsection{The Underdetermined System with more Unknowns than Equations}

If one poses the $l_p$ approximation problem in solving an underdetermined set of
equations (case 3 from Chapter 3), it comes from defining the solution norm as 
\begin{equation}
		||x||_p = \biggl( \sum_n |x(n)|^p \biggr)^{1/p}
\label{pn03}\end{equation}
and finding ${\bf x}$ to minimizing this p-norm while satisfying ${\bf Ax=b}$.

It has been shown this is equivalent to solving a least weighted norm problem
for specific weights.
\begin{equation}
		||x||_p = \biggl( \sum_n w(n)^2 |x(n)|^2 \biggr)^{1/2}
\label{pn03a}\end{equation}
The development follows the same arguments as in the previous section but using the formula 
\cite{ivan4,israel} derived in (\ref{lag3})
\begin{equation}
		{\bf x} ={\bf [W^T W]^{-1} A^T  \biggl[A  [W^T W]^{-1}  A^T \biggr]^{-1} b}
\label{pn33}\end{equation}
with the weights, $w(n)$, being the diagonal of the matrix, ${\bf W}$, 
in the iterative algorithm to give the minimum weighted solution norm in the same way as (\ref{pn3})
gives the minimum weighted equation error.

A Matlab program that implements these ideas applied to our pseudoinverse
problem with more unknowns than equations   (case 3a) is:
\begin{verbatim}
% m-file IRLS2.m to find the optimal solution to Ax=b 
%  minimizing the L_p norm ||x||_p, using IRLS.
%  Newton iterative update of solution, x, for  M < N.
%  For 2<p<infty, use homotopy parameter K = 1.01 to 2  
%  For 0<p<2, use K = approx 0.7 to 0.9
%  csb 10/20/2012
function x = IRLS2(A,b,p,K,KK)
if nargin < 5, KK= 10;  end; 
if nargin < 4, K = .8;  end;
if nargin < 3, p = 1.1; end;
pk = 2;                                 % Initial homotopy value
x  = pinv(A)*b;                         % Initial L_2 solution
E = [];
for k = 1:KK   
   if p >= 2, pk = min([p, K*pk]);      % Homotopy update of p
      else pk = max([p, K*pk]); end
   W  = diag(abs(x).^((2-pk)/2)+0.00001);  % norm weights for IRLS
   AW = A*W;                            % applying new weights
   x1 = W*AW'*((AW*AW')\b);             % Weighted L_2 solution
   q  = 1/(pk-1);                       % Newton's parameter
   if p >= 2, x = q*x1 + (1-q)*x; nn=p; % Newton's partial update for p>2
      else x = x1; nn=1; end            % no Newton's partial update for p<2
   ee = norm(x,nn);  E = [E ee];        % norm at each iteration
end;
plot(E)
\end{verbatim}
This approach is useful in sparse signal processing and for frame representation.

\section{The Chebyshev, Minimax, or $L_\infty$ Appriximation}

The {\bf Chebyshev} optimization problem minimizes the maximum error:
\begin{equation}
		\epsilon_m = \max_n |\epsilon (n)|
\label{em}\end{equation}
This is particularly important in filter design.  The Remez exchange algorithm
applied to filter design as the Parks-McClellan algorithm is very efficient \cite{csbcnxfd}.
An interesting result is the limit of an $||{\bf x}||_p$ optimization as 
$p \rightarrow \infty$ is the Chebyshev optimal solution.  So, the 
Chebyshev optimal, the minimax optimal, and the $L_\infty$ optimal are
all the same \cite{cheney,csbcnxfd}. 

A particularly powerful theorem which characterizes a solution to ${\bf Ax=b}$ is given
by Cheney \cite{cheney} in Chapter 2 of his book:

\begin{itemize}
\item {\bf A Characterization Theorem:} {\it For an $M$ by $N$ real matrix, ${\bf A}$ with $M > N$, every minimax solution 
${\bf x}$ is a minimax solution of an appropriate $N+1$ subsystem of the $M$ equations.  This
optimal minimax solution will have at least $N+1$ equal magnitude errors and they will be larger
than any of the errors of the other equations.}
\end{itemize}

This is a powerful statement saying an optimal minimax solution will have out of $M$, at least $N+1$ maximum
magnitude errors and they are the minimum size possible.  What this theorem doesn't state is which of the
$M$ equations are the $N+1$ appropriate ones.  Cheney develops an algorithm based on this
theorem which finds these equations and exactly calculates this optimal solution in a finite number 
of steps.  He shows how this
can be combined with the minimum $||{\bf e}||_p$ using a large $p$, to make an efficient solver for
a minimax or Chebyshev solution.

This theorem is similar to the Alternation Theorem \cite{csbcnxfd} but more general and, therefore,
somewhat more difficult to implement.

\section{The $L_1$ Approximation and Sparsity} 

The {\bf sparsity} optimization is to minimize the number of non-zero terms in a vector. 
A ``pseudonorm", $||{\bf x}||_0$, is sometimes used to denote a measure of sparsity. 
This is not convex, so is not really a norm but the convex (in the limit) norm
$||{\bf x}||_1$ is close enough to the $||{\bf x}||_0$ to give the same
sparsity of solution \cite{donoho71}.  Finding a sparse solution is not easy but interative 
reweighted least squares (IRLS) \cite{bbs,irlsvb}, weighted norms \cite{gorodnitsky,DaubSparse}, and
a somewhat recent result is called Basis Pursuit \cite{sschen,sschen2} are possibilities.

This approximation is often used with an underdetermined set of equations (Case 3a) to obtain
a sparse solution ${\bf x}$.

Using the IRLS algorithm to minimize the $l_p$ equation error often gives a sparse error if
one exists.  Using the algorithm in the illustrated Matlab program with $p=1.1$ on the problem in 
Cheney \cite{cheney} gives a zero error in equation 4 while using no larger $p$ gives any zeros.

\chapter{General Posing of Simultaneous Equations}\label{cc3}

Consider a mixture of the first and second problems posed in the introduction.
The first problem assumed that ${\bf x}$ is given and ${\bf b}$ is to be found.
The second problem assumed that ${\bf b}$ is given and ${\bf x}$ is to be found.
We now again consider
\be
    {\bf F X = Y}
\label{1a7}\ee
but now let there be $K$ unknowns in the vector ${\bf Y}$ and $N-K$ givens or 
known values.  The necessary balance of knowns and unknowns is maintained
by having $N-K$ unknowns in the vector ${\bf X}$ and $K$ givens or 
known values.  With this generalization of the problem, we still have the same
number of equations and unknowns but the unknowns are now divided to be
on both sides of the equation.  Note that this has the original problem-one as
a special case for $K=0$ and problem-two results from $K=N$ but now allows a
much more versatile mixture of the two.  The integer $K$ with $0\leq K \leq N$
is the parameter that determines the mixture.

To show that we can solve for the unknowns, we now re-order the rows of 
${\bf F}$ in (\ref{1a7}) so that the unknowns in ${\bf Y}$ 
are the first  $K$ entries.  We then re-order the columns so that the givens (knowns)  
in ${\bf X}$ are the first $K$ entries.  These reordered equations are partitioned 
in the form of
\be
\left[ \begin{array}{cc}
        {\bf A} & {\bf B} \\
        {\bf C} & {\bf D} 
        \end{array} \right]
            \left[ \begin{array}{c}
            {\bf X_1}\\
            {\bf X_2}
                  \end{array} \right]
  =
    \left[ \begin{array}{c}
    {\bf Y_1}\\
    {\bf Y_2}
         \end{array} \right]
\label{1a8}\end{equation}
with the matrix ${\bf A}$ being $K$ by $K$ and ${\bf D}$ being $N-K$ by $N-K$ 
and with ${\bf X_1}$ and  ${\bf Y_2}$ being given and  ${\bf X_2}$ and ${\bf Y_1}$ 
being unknown.

Eliminating ${\bf X_2}$ gives for ${\bf Y_1}$
	
\be
	[{\bf A} - {\bf B D}^{-1}{\bf C}] {\bf X_1} + [{\bf B D}^{-1}] {\bf Y_2} = {\bf Y_1}
\label{1a9}\ee
which requires ${\bf X_2}$ to be 
\be
	-{\bf D}^{-1}{\bf C} {\bf X_1} + {\bf D}^{-1} {\bf Y_2} = {\bf X_2}
\label{1a10}\ee

Equations (\ref{1a9}) and (\ref{1a10}) can be written in one partitioned matrix equation as
\be
\left[ \begin{array}{cc}
        {\bf A} - {\bf B D}^{-1}{\bf C} &{\bf B D}^{-1} \\
        -\bf{ D}^{-1}{\bf C} &  {\bf D}^{-1} 
        \end{array} \right]
            \left[ \begin{array}{c}
            {\bf X_1}\\
            {\bf Y_2}
                  \end{array} \right]
  =
    \left[ \begin{array}{c}
    {\bf Y_1}\\
    {\bf X_2}
         \end{array} \right]
\label{1a11}\end{equation}
which is in the original form of having the $N$ knowns on the left hand side of the equation
and the $N$ unknowns (to be calculated) on the right hand side.

Note that the original problem-1, which is the case for $K=0$, causes (\ref{1a9}) to become
simply ${\bf A X_1} = {\bf Y_1}$ and problem-2 with $K=N$ gives ${\bf D^{-1} Y_2 = X_2}$.

This mixed formulation of simultaneous equations allows a linear algebra description of
IIR (Infinite duration Impulse Response) digital filters and the use of partitions allows linear
design of IIR filters which interpolate samples of a desired frequency response \cite{csbcnxfd}.  With
the definition of an equation error, it also allows the optimal least squared equation error
approximation.  This is used in implementing Prony's method, Pade's method, and linear
prediction \cite{csbPPLP}.

Note also that this mixed formulation can also be stated in an over determined or under
determined form which will require approximation in finding an optimal solution,
see chapter \ref{problem}.

The general posing of simultaneous equations also gives some interesting insights into sampling
theory.  For example, if the matirx ${\bf F}$ is the discrete Fourier transform (DFT) matrix, and
the signal ${\bf X}$ is band limited, then (\ref{1a11}) describes the case with ${\bf Y_2}= {\bf 0}$
implying that the total signal ${\bf X}$ can be calculated from $K$ samples in ${\bf X_1}$ by finding
${\bf X_2}$ from (\ref{1a10}).  This can be viewed as a generalized sampling theorem.

A particularly interesting and important application of this formulation is the calculation of
the sparse FFT (fast Fourier transform) \cite{hassanieh,pawar,hsieh}.  In this problem, it is 
known that the signal has only $K$ non-zero spectral values (usually with $K<<N$).  
In other words, from the specifics of the physical 
problem, it is known that $N-K$ values of the DFT (discrete Fourier transform) of the signal
are zero.  From (\ref{1a9}), we see that the sparseness requires ${\bf Y_2}={\bf 0}$ and
the matrix ${\b F}$ in (\ref{1a7}) is the DFT matrix.  The desired DFT values are ${\bf Y_1}$
and are given by
\be
	[{\bf A} - {\bf B D}^{-1}{\bf C}] {\bf X_1}  = {\bf Y_1}
\label{1a12}\ee
which can be calculated from any $K$ samples of the $K$-sparse signal ${\bf X}$ requiring
at most $O(K^2)$ operations.  ${\bf D}$ is non-singular if ${\bf F}$ is orthogonal (which the
DFT matrix is) and may be in other cases.

From this formulation, it is seen that a length-$N$ $K$-sparse signal lies in a $K$ dimensional
subspace of the $N$ dimensional signal space.  The DFT of any signal in the subspace can be
calculated from $K$ samples of the signal using (\ref{1a12}).  Most of the recent work on sparse
FFTs does not assume the location of the $K$ non-zero terms in ${\bf Y}$ is known ahead of time
but are discovered as part of the solution.  This seems to require more samples as well as more
computation.  The most common approach ``filters'' the signal and finds the solution in ``bins''
in the frequency domain \cite{hassanieh,hassanieh2}. The number of operations for these
approachs seems to be $O(K \log(N))$.  Another approach uses the Chinese remainder
theorem \cite{pawar} to determine samples of ${\bf X}$ or orthogonal polynomials \cite{hsieh}.
These claim a number of operations of $O(K \log(K))$.

\chapter{Constructing the Operator}

Solving the third problem posed in the introduction to these notes is
rather different from the other two.  Here we want to find an operator or
matrix that when multiplied by \x gives \bb.  Clearly a solution to this
problem would not be unique as stated.  In order to pose a better defined
problem, we generally give a set or family of inputs \x and the
corresponding outputs \bb.  If these families are independent, and if the
number of them is the same as the size of the matrix, a unique matrix is
defined and can be found by solving simultaneous equations.  If a smaller
number is given, the remaining degrees of freedom can be used to satisfy
some other criterion.  If a larger number is given, there is probably no
exact solution and some approximation will be necessary. 

If the unknown operator matrix is of dimension $M$ by $N$, then we take $N$ inputs
${\bf x_k}$ for $k = 1, 2, \cdots, N$, each of dimension $N$ and the 
corresponding $N$ outputs ${\bf b_k}$, each of dimension $M$ and form
the matrix equation:
\begin{equation}
			{\bf AX = B}
\label{r0}\end{equation}
where ${\bf A}$ is the $M$ by $N$ unknown operator, ${\bf X}$ is the $N$ by $N$
input matrix with $N$ columns which are the inputs ${\bf x_k}$ and ${\bf B}$
is the $M$ by $N$ output matrix with columns ${\bf b_k}$.  The operator matrix 
is then determined by:
\begin{equation}
			{\bf A = B X^{-1}}
\label{r00}\end{equation}
if the inputs are independent which means ${\bf X}$ is nonsingular.

This problem can be posed so that there are more (perhaps many more)
inputs and outputs than $N$ with a resulting equation error which can
be minimized with some form of pseudoinverse.

Linear regression can be put in this form.  If our matrix equation is
\be
    {\bf A x = b}
\label{r1}\ee
where ${\bf A}$ is a row vector of unknown weights and ${\bf x}$ is a
column vector of known inputs, then ${\bf b}$ is a scaler inter product. 
If a seond experiment gives a second scaler inner product from a second
column vector of known inputs, then we augment ${\bf X}$ to have two
rows and ${\bf b}$ to be a length-2 row vector.  This is continued for
$N$ experiment to give (\ref{r1}) as a 1 by $N$ row vector times an $M$ by $N$ matrix
which equals a 1 by $M$ row vector.  It this equation is transposed, it is in the form
of (\ref{r1}) which can be approximately solved by the pesuedo inverse to 
give the unknown weights for the regression.

Alternatively, the matrix may be constrained by structure to have less
than $N^2$ degrees of freedom.  It may be a cyclic convolution, a non
cyclic convolution, a Toeplitz, a Hankel, or a Toeplitz plus Hankel
matrix.

A problem of this sort came up in research on designing efficient prime length fast
Fourier transform (FFT) algorithms where \x is the data and \bb is the FFT
of \x.  The problem was to derive an operator that would make this
calculation using the least amount of arithmetic.  We solved it using a
special formulation \cite{J&B:85} and Matlab.

This section is unfinished.

\chapter{Topics that might be Added}

The following topics may added to these notes over time:

\ben 
\item Different norms on equation error and solution error and solution size
\item exchange algorithm for Cheby approx from Cheney plus a program
\item Freq. sampling \cite{yonina}, least squares, Chebyshev design of FIR filters
\item Block formulation of FIR and IIR digital filters, Prony and Pade approximation
\item Periodically time varying filters, Two-frequency formulation
\item State variable modelling of dynamic systems; feedback control systems, sol by diagonalizing ${\bf A}$
\item Regression, Machine Learning, Neural Networks (layers of linear and non-linear operators)
\item The eigenvalue problem, eigenvector expansions, diagonalization of a matrix, and singular value decomposition (SVD)
\item Quadratic forms, other optimizaion measures e.g. polynomials
\item Other error definitions for approximate solutions. Sparsity \cite{IvanSparse} and ${l_1}$ approximation.
\item Use of Matlab, Octave, SciLab, Mathematica, Maple, LabVIEW, R, Python, etc.
\item Constrained approximation, Kuhn-Tucker. 
\item Expansion and completion of Chapter 6 on Constructing the Operator.
\een
Please contact the author at [csb@rice.edu] with any errors, recommendations, or comments. Anyone is free to use
these note in any way as long as attribution is given according to the Creative Commons (cc-by) rules.

\markboth{Bibliography}{Bibliography} \def\CHHEAD{Bibliography}
\bibliographystyle{unsrt}
\addcontentsline{toc}{chapter}{Bibliography}
{\small
\bibliography{art,fft,wave,burrus,book,filter,burrus3}
}

\end{document}